\documentclass[12pt,fleqn]{article}
\usepackage{graphicx}

\usepackage{latexsym}

\usepackage{amsmath}
\usepackage{amsthm}
\usepackage{amssymb}
\usepackage{amsfonts}

\numberwithin{equation}{section}

\begin{document}

\newcommand{\rf}[1]{(\ref{#1})}
\newcommand{\rff}[2]{(\ref{#1}\ref{#2})}

\newcommand{\ba}{\begin{array}}
\newcommand{\ea}{\end{array}}

\newcommand{\be}{\begin{equation}}
\newcommand{\ee}{\end{equation}}

\newcommand{\const}{{\rm const}}
\newcommand{\ep}{\varepsilon}
\newcommand{\Cl}{{\cal C}}
\newcommand{\rr}{{\vec r}}
\newcommand{\ph}{\varphi}
\newcommand{\R}{{\mathbb R}}  
\newcommand{\N}{{\mathbb N}}
\newcommand{\Z}{{\mathbb Z}}
\newcommand{\C}{{\mathbb C}}  

\newcommand{\e}{{\bf e}}

\newcommand{\m}{\left( \ba{r}}
\newcommand{\ema}{\ea \right)}
\newcommand{\mm}{\left( \ba{cc}}
\newcommand{\miv}{\left( \ba{cccc}}

\newcommand{\scal}[2]{\mbox{$\langle #1 \! \mid #2 \rangle $}}
\newcommand{\ods}{\par \vspace{0.3cm} \par}
\newcommand{\dis}{\displaystyle }
\newcommand{\mc}{\multicolumn}
\newcommand{\no}{\par \noindent}

\newcommand{\sinc}{ {\rm sinc\,} }
\newcommand{\tanhc}{{\rm tanhc} }
\newcommand{\tanc}{{\rm tanc} }

\newtheorem{prop}{Proposition}[section]
\newtheorem{Th}[prop]{Theorem}
\newtheorem{lem}[prop]{Lemma}
\newtheorem{rem}[prop]{Remark}
\newtheorem{cor}[prop]{Corollary}
\newtheorem{Def}[prop]{Definition}
\newtheorem{open}{Open problem}
\newtheorem{ex}[prop]{Example}
\newtheorem{exer}[prop]{Exercise}

\newenvironment{Proof}{\par \vspace{2ex} \par
\noindent \small {\it Proof:}}{\hfill $\Box$ 
\vspace{2ex} \par }

\title{\bf 
Locally exact modifications of discrete gradient schemes}
\author{
 {\bf Jan L.\ Cie\'sli\'nski}\thanks{\footnotesize
 e-mail: \tt janek\,@\,alpha.uwb.edu.pl} 
\\ {\footnotesize Uniwersytet w Bia{\l}ymstoku,
Wydzia{\l} Fizyki}
\\ {\footnotesize ul.\ Lipowa 41, 15-424
Bia{\l}ystok, Poland}
}

\date{}

\maketitle

\begin{abstract}
Locally exact integrators preserve linearization of the original system at every point. We  construct energy-preserving locally exact discrete gradient schemes for arbitrary multidimensional canonical Hamiltonian systems by modifying classical discrete gradient schemes. Modifications of this kind are found for  any discrete gradient. 
\end{abstract}

\ods




\section{Introduction}

In this Letter we  introduce new classes of energy-preserving numerical methods for multidimensional canonical Hamiltonian systems by a modification of classical discrete gradient schemes in a locally exact way.  Conservative integrators  have been used for many years for simulating dynamics of systems of particles  \cite{Gree,LaG}. 
More recently discrete gradient methods have been developed in the context of geometric numerical integration \cite{MQ}, see \cite{IA,Gon}.  In particular, numerical integrators constructed by Quispel and his co-workers preserve integrals of motion of a given system of ordinary differential equations \cite{QC,QT,MQR1}. It is well known that preservation of geometric properties by numerical algorithms is of considerable advantage \cite{HLW}. 

A numerical scheme is called locally exact if its linearization coincides with the exact discretization of linearized equations \cite{CR-PRE,CR-BIT,Ci-locex}.  We point out that exact discretization gives the exact solution at each time step. Thus two known procedures are combined:  approximation of nonlinear systems by linear equations and explicit exact discretizations of linear ordinary differential equations with constant coefficients. 
Exact discretization of the linearized equation (i.e., the exponential Euler method) has been considered a long time ago \cite{Pope}, compare \cite{Law}.  Recently, a similar notion (preservation of the linearization at all fixed points) appeared in \cite{MQTse}, see also \cite{CR-long}. Locally exact numerical schemes can be considered as a special case of exponential integrators \cite{MW,HoL1,CCO}. 
Our approach considers non-standard modifications  of  numerical schemes (see \cite{Mic}), parameterized by some functions (e.g., a matrix denoted by $\pmb{\delta}$). In contrast to the original Mickens approach, where such functions are postulated or guessed, we determine them  by requiring local exactness.   The final formulae are sometimes reminiscent of results generated by Gautschi-type methods \cite{Gau,HoL2}. 

Locally exact modifications of discrete gradient methods  for Hamiltonian systems with one degree of freedom were studied  in \cite{CR-PRE,CR-BIT}.  We modified the discrete gradient scheme in a locally exact way preserving its main geometric property: the exact conservation of the energy integral. Numerical experiments  show that locally exact modifications can increase the accuracy by several orders of magnitude. The multidimensional case is much more complicated.  First results, confined to the coordinate increment discrete gradient and its symmetric modification,  were reported in \cite{Ci-locex}.  In this paper we present general results that are valid for any discrete gradient.

\section{Local exactness}
\label{sec-locex}

We consider an ordinary differential equation (ODE) \ $\dot {\pmb y} = F ({\pmb y})$ \ with solution ${\pmb y} (t) \in \R^m$ (for some $m \in \N$ and  initial condition ${\pmb y} (t_0) = {\pmb y}_0$), and a difference equation with variable time step $h_n$ and solution $({\pmb y}_n)\subset \R^m$. We denote $h_n = t_{n+1} - t_n$, hence we get a sequence $t_n$ for $n \in \N \cup \{0\}$.   In other words,  there is a map that takes the sequence $(t_n)$ to the sequence $(\pmb{y}_n)$ and each vector $\pmb{y}_n$ is the approximate solution at time $t_n$. We say that  the difference equation 
 is an  exact discretization of  the  ODE  if \ ${\pmb y}_n = {\pmb y} (t_n)$ for $n\in \N \cup \{0\}$.

It is well known that any linear ODE with constant coefficients admits the exact discretization in an explicit form \cite{Potts}, see also \cite{Mic,Ag,CR-ade}. 
Moreover, exact discretizations were applied in the numerical solution of the classical Kepler problem \cite{Ci-Kep,Ci-Koz,Ci-oscyl} and the wave equation \cite{Ci-oscyl}.   
The central topic of our paper is another fruitful direction of using exact integrators,  namely the so called locally exact discretizations \cite{CR-PRE,Ci-oscyl}. 
Motivated by the results of \cite{CR-PRE,CR-BIT} we propose the following definition (see also \cite{Ci-locex}). 

\begin{Def}  \label{def-locex}
A numerical scheme ${\pmb y}_{n+1} = \Phi ({\pmb y}_n, h_n)$ for an autonomous ordinary differential equation $\dot {\pmb y} = F ({\pmb y})$  is {\it locally exact} if there exist a sequence $(\pmb{\bar y}_n)$ such that $\pmb{\bar y}_n  - {\pmb y}_n = O (h_n)$ and the linearization of the scheme 
around $\pmb{\bar y}_n$ is identical with the exact discretization of the differential equation linearized around $\pmb{\bar y}_n$ (for any $n$). 
\end{Def}

In particular cases we usually assume \ $\pmb{\bar y}_n = \pmb{y}_n$ \ or \  $\pmb{\bar y}_n = \frac{1}{2} \left( \pmb{y}_n + \pmb{y}_{n+1} \right)$, see \cite{CR-PRE,CR-BIT,Ci-locex}. It is worthwhile to point out that all proofs included in this Letter are valid for any choice of $(\pmb{\bar y}_n)$.

\ods

Similar concept (``linearization-preserving'' schemes)  has been recently formulated in \cite{MQTse}. A scheme is said to be linearization-preserving if it is linearization preserving at all fixed points. All locally exact schemes are linearization-preserving but, in general, a linearization-preserving scheme may not be locally exact. 

\ods

If ${\pmb y}$ is near  $\pmb{\bar y}_n$, then the equation $\dot {\pmb y} = F ({\pmb y})$   can be approximated by
\be  \label{approx1}
\dot {\pmb \nu} = F' (\pmb{\bar y}_n) {\pmb \nu}  +  F (\pmb{\bar y}_n) 
\ee
where $F'$ is the Fr\'echet derivative (Jacobi matrix) of $F$ and 
\be  \label{xi}
 {\pmb \nu} = {\pmb y} - \pmb{\bar y}_n \ . 
\ee
The exact discretization of the approximated equation \rf{approx1} is given by
\be  \label{exact-approx}
{\pmb \nu}_{n+1} = e^{ h_n F' (\pmb{\bar y}_n)} {\pmb \nu}_n + h_n \varphi_1 (h_n F' (\pmb{\bar y}_n) )  F (\pmb{\bar y}_n) \   
\ee
(compare  \cite{Ci-locex,Pope}), where $\varphi_1$ is an analytic function defined  by
\[
\varphi_1 (z) = \sum_{k=1}^\infty \frac{z^{k-1}}{k!} \  , 
\]
see, e.g., \cite{NW}. For $z \neq 0$ we have  $\varphi_1 (z) = (e^z - 1)/z$.  Moreover, 
\be  \label{fi1}
 (\varphi_1 (z))^{-1} = 1 - \frac{z}{2} - \sum_{k=1}^\infty  \frac{(-1)^{k} B_k z^{2k}}{(2 k)!} \ ,
\ee
where $B_k$ are Bernoulli numbers ($B_1=\frac{1}{6}$, $B_2=\frac{1}{30}$, $B_3=\frac{1}{42}$, etc.), compare e.g. \cite{Dw}. This series converges  for $|z| < 2 \pi$. 
Note that
\be  \label{exp-fi}
e^{h_n F' (\pmb{\bar y}_n)} = I + h_n F' (\pmb{\bar y}_n)  \varphi_1 (h_n F' (\pmb{\bar y}_n))  , 
\ee
which holds also for singular $F' (\pmb{\bar y}_n)$. In this Letter we do not need to assume  \ $\det F' (\pmb{\bar y}_n) \neq 0$.  \   

Here and in the next sections we often use analytic functions of matrices, see e.g. \cite{Ru,Is}. If  $g: \C \rightarrow \C$ is analytic on the disc $\{ z\in \C: |z|< r\}$ for some $r$, then it has a locally defined Taylor series  $g (z) = \sum_{n=0}^\infty g_n z^n$  and we can define $g (A)$ by 
\be  \label{ser}
  g (A) = \sum_{n=0}^\infty g_n A^n \ ,   
\ee
for any matrix $A$  (or a continuous linear operator in a Banach space)  such that  $\| A \| \leqslant r$, where  $\|\cdot\|$ is a  norm satisfying  $\| A B \| \leqslant \| A \| \|B \|$ for any operators  $A$ and $B$.     
If all eigenvalues $\lambda_1,\ldots,\lambda_m$ of a matrix $A$ are disctinct, then  $A$  has the spectral factorization  $A = T \text{diag} ( \lambda_1,\ldots, \lambda_m) T^{-1}$, where $k$th column of the matrix  $T$ is the eigenvector associated with $\lambda_k$,  and we can use a  compact  formula  $g (A) = T \text{diag} (  g (\lambda_1), \ldots, g(\lambda_m) ) T^{-1}$.

Locally exact discretizations have some qualitative advantages which follow directly from their definition. Namely, they preserve all fixed points of the considered differential equation and they are linearly stable and $A$-stable. Indeed, locally exact schemes applied to linear equations produce exact solutions and exact trajectories.

\section{Discrete gradients} 
\label{sec-locex-multi}

We consider arbitrary multidimensional Hamiltonian systems in canonical coordinates:
\be \label{multiham}
  {\dot x}^k = \frac{\partial H}{\partial p^k} \ , \quad 
{\dot p}^k = - \frac{\partial H}{\partial x^k} \ , \quad (k = 1,\ldots,m) \ , 
\ee
where $H = H(\pmb{x}, \pmb{p})$ and 
$\pmb{x}:= (x^1,\ldots, x^m)$, $\pmb{p}:= (p^1,\ldots, p^m)$. 
Denoting $\pmb{y} := (\pmb{x}, \pmb{p})^T \in \R^{2m}$ we obtain a more concise form of 
the Hamiltonian system:  
\be  \label{F-ham} 
  \pmb{\dot y} =   F (\pmb{y}) \ , \quad  F (\pmb{y}) = S \nabla H \ , \quad 
S = \mm 0 & I \\ - I & 0 \ema  \ .
\ee 
where 
\be  \label{ozn-grad}
\nabla H \equiv  \frac{\partial H}{\partial \pmb{y}} \equiv  H_{\pmb{y}} := (H_{y^1}, H_{y^2},\ldots, H_{y^{2 m}})^T \ , \qquad  
H_{y^j} = \frac{\partial H}{\partial y^j} \ . 
\ee
It is well known that Hamiltonian $H$ is a first integral of the system \rf{F-ham} since 
\be \label{H=const} 
\frac{d H}{d t} = \sum_{k=1}^{2m} \frac{\partial H}{\partial y^k} \ \frac{d   y^k}{d t}  
\equiv \scal{ \nabla H}{\pmb{\dot y} }  = \scal{ \nabla H}{ S \nabla H } = 0 \ , 
\ee
where the bracket (denoting a scalar product) is defined by the second equality. The last equality  of 
\rf{H=const} holds for any skew-symmetric matrix $S$.

\begin{Def} A discrete gradient
\be
 {\bar \nabla} H   = \left( \frac{ {\Delta} H}{\Delta y^1}, \frac{ {\Delta} H}{\Delta y^2}, \ldots,\frac{ {\Delta} H}{\Delta y^{2 m}}  \right) 
\ee
is a continuous  function ${\bar \nabla} H: \R^{2m} \times \R^{2m} \ni (\pmb{y}_n, \pmb{y}_{n+1}) \mapsto  
\bar \nabla H (\pmb{y}_n, \pmb{y}_{n+1}) \in \R^{2 m}$,  
such that \cite{Gon,MQR1}: 
\be  \label{disgrad-def}
\scal{{\bar \nabla} H (\pmb{y}_n, \pmb{y}_{n+1}) }{\pmb{y}_{n+1} - \pmb{y}_n} = H (\pmb{y}_{n+1}) - H (\pmb{y}_n) \ , 
\ee 
\be \label{disgrad-lim-def}
\lim_{\pmb{\tilde y} \rightarrow \pmb{y}} {\bar \nabla} H (\pmb{y}, \pmb{\tilde y}) = \nabla H (\pmb{y})\ . 
\ee
\end{Def}

\ods

Discrete gradients are non-unique, compare \cite{IA,Gon,MQR1}. Many functions satisfy these conditions, for example 
the so called coordinate increment discrete gradient \cite{IA}: 
\be \ba{l} \dis \label{grad-incre} 
 \frac{\Delta H}{\Delta y^1} = \frac{H(y_{n+1}^1, y_n^2, y_n^3,\ldots, y_n^{2 m}) - H (y_n^1, y_n^2, y_n^3,\ldots, y_n^{2 m})}{y_{n+1}^1 - y_n^1}   \\[3ex]\dis  
 \frac{\Delta H}{\Delta y^2} = \frac{H(y_{n+1}^1, y_{n+1}^2, y_n^3,\ldots, y_n^{2 m}) - H (y_{n+1}^1, y_n^2, \ldots, y_n^{2 m})}{y_{n+1}^2 - y_n^2}   \\[2ex]\dis  
\dotfill  \  \\[1ex]\dis  
\frac{\Delta H}{\Delta y^{2m}} = \frac{H(y_{n+1}^1, y_{n+1}^2, \ldots, y_{n+1}^{2m}) - H (y_{n+1}^1, y_{n+1}^2, \ldots, y_n^{2m}) }{y_{n+1}^{2m} - y_n^{2m}} \ .
\ea \ee

In fact, any permutation of $2 m$ coordinates $x^k, p^j$ can be identified with components of $\pmb{y}$.   
Thus we obtain $(2 m)!$ discrete gradients of this type (some of them may happen to be identical). 

Having any discrete gradient we can define the related symmetric discrete gradient
\be  \label{grad-sym}
 {\bar \nabla}_s H (\pmb{y}_n, \pmb{y}_{n+1}) = \frac{1}{2} \left( {\bar \nabla} H (\pmb{y}_n, \pmb{y}_{n+1}) + {\bar \nabla} H (\pmb{y}_{n+1}, \pmb{y}_n) \right) \ .
\ee
One can easily verify that ${\bar \nabla}_s H$ satisfies conditions \rf{disgrad-def}, \rf{disgrad-lim-def} provided that they are satisfied by $\bar \nabla H$.

In order to construct locally exact modifications we need to linearize  discrete gradients defined by \rf{grad-incre} and \rf{grad-sym}. 

\ods

\begin{Def} \label{def-AB}
We define $2 m \times 2 m$ matrices $A, B$ as follows
\be \ba{l} \dis  \label{AB}
A = \left. \frac{\partial \bar{\nabla}H (\pmb{\bar y}, \pmb{  y}) }{\partial \pmb{  y}} \right|_{\footnotesize \ba{l} \pmb{\bar  y} = \pmb{\bar y}_n \\ \pmb{  y} = \pmb{\bar y}_n \ea} \ , \quad  
 \left. B = \frac{\partial \bar{\nabla}H (\pmb{\bar y}, \pmb{  y}) }{\partial \pmb{\bar y}  } \right|_{\footnotesize \ba{l} \pmb{\bar  y} = \pmb{\bar y}_n \\ \pmb{  y} = \pmb{\bar y}_n \ea}   \  .
\ea \ee
Thus linearization of the discrete gradient around $\pmb{\bar y}_n$ is given by
\be  \label{Tay-AB} 
  \bar \nabla H (\pmb{y}_n, \pmb{y}_{n+1}) \approx H_{\pmb{y}} +  A \pmb{\nu}_{n+1} + B \pmb{\nu}_n  \ , 
\ee
where we denoted $\pmb{\nu}_n = \pmb{y}_n - \pmb{\bar y}_n$ \ and \ $H_{\pmb{y}}$ is evaluated at \ $\pmb{\bar y}_n$. 
\end{Def}

\ods

In the case of the coordinate increment discrete gradient \rf{grad-incre} entries of matrices  $A, B$ can be obtained by straightforward calculation (compare \cite{Ci-locex}):
\be
  A_{jk} = \left\{  \ba{cl}  H_{y^j y^k} & \  (j > k) \\ 
\frac{1}{2} H_{y^k y^k} & \    (j=k)  \\ 0 & \   (j < k)   \ea \right.   , \qquad
  B_{jk} = \left\{  \ba{cl}  0 & \  (j > k) \\ 
\frac{1}{2} H_{y^k y^k} & \   (j=k)  \\ H_{y^j y^k}  & \ (j < k)   \ea \right.   .
\ee

\ods

\begin{Th}  \label{Th-AB}
For any discrete gradient we have:
\be \ba{l} \label{A+B}
B = A^T \ , \quad A + B = H_{\pmb{y} \pmb{y} }\ , \\[2ex] \dis 
A^T + A = H_{\pmb{y} \pmb{y} } \ , \quad B^T + B = H_{\pmb{y} \pmb{y} } \ , 
\ea \ee
where $H_{\pmb{y} \pmb{y}} $ is the Hessian matrix of $H$, evaluated at $\pmb{\bar y}_n$.  
\end{Th}

\begin{Proof}
We expand \rf{disgrad-def} in a Taylor series, expanding $\pmb{y}_n$ and $\pmb{y}_{n+1}$  around $\pmb{\bar y}_n$. Taking into account \rf{Tay-AB}, 
\be
H (\pmb{y}_n) =  H  + \scal{H_{\pmb{y}}}{\pmb{\nu}_n} + \frac{1}{2} \scal{\pmb{\nu}_n}{H_{\pmb{y}\pmb{y}} \pmb{\nu}_n } + \ldots \ ,
\ee
and the analogical expansion for $H (\pmb{y}_{n+1})$, we observe that the linear parts of both sides of \rf{disgrad-def} are obviously identical. Considering quadratic terms we get the following equations: 
\be 
\scal{\pmb{\nu}_{n+1}}{ A \pmb{\nu}_{n+1} } = \frac{1}{2} \scal{\pmb{\nu}_{n+1}}{ H_{\pmb{y}\pmb{y}} \pmb{\nu}_{n+1}}  , 
\quad 
\scal{\pmb{\nu}_n}{ B \pmb{\nu}_n} = \frac{1}{2} \scal{\pmb{\nu}_n}{ H_{\pmb{y}\pmb{y}} \pmb{\nu}_n}  ,  
\ee
\be
\scal{ \pmb{\nu}_{n+1} }{ B \pmb{\nu}_n } - \scal{ A \pmb{\nu}_{n+1} }{ \pmb{\nu}_n } = 0 \ ,
\ee
which have to be satisfied for any $\pmb{\nu}_n$, $\pmb{\nu}_{n+1}$. After elementary algebraic manipulations we obtain:
\be \label{1122}
\scal{\pmb{\nu}_{n+1}}{ \left( A  - \frac{1}{2} H_{\pmb{y}\pmb{y}} \right) \pmb{\nu}_{n+1} } = 0 \  , 
\quad 
\scal{\pmb{\nu}_n}{ \left( B  - \frac{1}{2} H_{\pmb{y}\pmb{y}} \right) \pmb{\nu}_n } = 0 \  , 
\ee
\be  \label{33}
\scal{ \pmb{\nu}_{n+1} }{ (B - A^T) \ \pmb{\nu}_n } = 0 \ .
\ee
Equations \rf{1122} imply that matrices \ $A  - \frac{1}{2} H_{\pmb{y}\pmb{y}}$ \ and \ $B  - \frac{1}{2} H_{\pmb{y}\pmb{y}}$ \ are skew-symmetric. Hence $A^T + A = H_{\pmb{y} \pmb{y} } = B^T + B$. Finally, from \rf{33} we get $B = A^T$ and, as a consequence, $A + B = H_{\pmb{y} \pmb{y} }$. 
\end{Proof}

\begin{cor} \label{cor-sym}
Linearization of any symmetric discrete gradient is given by
\be
\bar \nabla_s H (\pmb{y}_n, \pmb{y}_{n+1}) \approx  H_{\pmb{y}} + \frac{1}{2} H_{\pmb{y} \pmb{y}} \left( \pmb{\nu}_n + \pmb{\nu}_{n+1} \right) \ . 
\ee
\end{cor}

Indeed, in the symmetric case  $B = A$. Therefore we can simplify \rf{Tay-AB} using $A + B = H_{\pmb{y} \pmb{y} }$.

\section{Modified discrete gradient schemes}

The standard discrete gradient scheme is given by 
\be
 \pmb{y}_{n+1} - \pmb{y}_n = h_n S {\bar \nabla} H  \ ,
\ee
where ${\bar \nabla} H = {\bar \nabla} H(\pmb{y}_n, \pmb{y}_{n+1})$ is a given discrete gradient. 
Replacing $S$ by a skew symmetric matrix, we obtain a large class of energy-preserving numerical schemes, closely related to the discrete gradient method.

\begin{lem}   \label{lem-zachow} 
Let $\Lambda$ be  \ $2 m \times 2 m$ \ matrix (depending on $h_n, \pmb{y}_n$ and $\pmb{y}_{n+1}$) such that
\be   \label{AS} 
 \Lambda^T = - \Lambda \ , \qquad \lim_{h_n \rightarrow 0} \frac{1}{h_n} \Lambda = S \ , \qquad S = \mm 0 & I \\ - I & 0 \ema 
\ee
and ${\bar \nabla} H =  {\bar \nabla} H(\pmb{y}_n, \pmb{y}_{n+1})$ is any discrete gradient. Then, the numerical scheme  
\be    \label{GR-LAM}
  \pmb{y}_{n+1} - \pmb{y}_n = \Lambda {\bar \nabla} H (\pmb{y}_n, \pmb{y}_{n+1}) \ ,
\ee
is a consistent  energy-preserving integrator for Hamiltonian system \rf{F-ham}. 
\end{lem}

\begin{Proof}
The consistency follows immediately from \rf{AS}. The energy preservation can be shown in the standard way. Using the scalar product, we multiply both sides of \rf{GR-LAM} by ${\bar \nabla} H$
\be
\scal{{\bar \nabla} H}{ \pmb{y}_{n+1} - \pmb{y}_n } = \scal{{\bar \nabla} H}{ \Lambda {\bar \nabla} H } \ .
\ee
By \rf{disgrad-def} the left-hand side equals $H (\pmb{y}_{n+1}) - H (\pmb{y}_n)$. The right hand side vanishes due to the skew symmetry of $\Lambda$. Hence $H (\pmb{y}_{n+1}) = H (\pmb{y}_n)$. 
\end{Proof}

The following theorem characterizes locally exact modifications of the discrete gradient method applied to equation \rf{F-ham}. \ Any \ $(\pmb{\bar y}_n)$ \ satisfying \  $\pmb{\bar y}_n - \pmb{y}_n = O (h_n)$ \ is acceptable.

\begin{Th}  \label{Th-Lam-locex}
Numerical scheme \ $ \pmb{y}_{n+1} - \pmb{y}_n = \Lambda {\bar \nabla} H$ (where  $\Lambda$ depends on $\pmb{\bar y}_n$ and $h_n$)   
is locally exact if
\be \label{lam} 
  \Lambda =   h_n  \Phi_1 S  \left(I + h_n  A \Phi_1 S \right)^{-1}  , 
\ee
where $\Phi_1 = \varphi_1 (h_n F')$,\ $F' = S H_{\pmb{y} \pmb{y}}$ (evaluated at $\pmb{\bar y}_n$),  
$A$ is given by \rf{AB} and we assume that  $\left(I + h_n  A \Phi_1 S \right)^{-1}$ exists.  
\end{Th}

\begin{Proof}  By virtue of \rf{Tay-AB} linearization of \rf{GR-LAM} (around $\pmb{\bar y}_n$)  is given by
\be  \label{lin2}
 \pmb{\nu}_{n+1} - \pmb{\nu}_n = \Lambda  ( A \pmb{\nu}_{n+1} + B \pmb{\nu}_n ) + \Lambda  H_{\pmb{y}} \ .
\ee
Hence, taking into account  $S H_{\pmb{y}} = F$ (where $F=F(\pmb{\bar y}_n)$),   we have
\be \label{nu-dis2} 
 \left( I -  \Lambda A \right) \pmb{\nu}_{n+1} =  \left( I +  \Lambda B \right) \pmb{\nu}_n + \Lambda S^{-1}  F \ .
\ee
The scheme \rf{GR-LAM} is locally exact if and only if \rf{nu-dis2} coincides with \rf{exact-approx}. Therefore, we require that 
\be   \label{11-eq}  
\left(I - \Lambda A \right) e^{h_n F'} =  I +  \Lambda  B  \ ,   
\ee
\be  \label{1-eq}
h_n \left( I -  \Lambda A \right)  \Phi_1   F = \Lambda S^{-1} F \ .
\ee
Using Theorem~\ref{Th-AB} (namely, $A+B=S^{-1} F'$) and equation \rf{exp-fi}, we transform equation \rf{11-eq} into  
\be   \label{2-eq}  
h_n \left( I -  \Lambda A \right)  \Phi_1   F' = \Lambda S^{-1} F' \ .
\ee
Both equations \rf{1-eq} and \rf{2-eq} are simultaneously satisifed if
\be
h_n \left( I -  \Lambda A \right)  \Phi_1  = \Lambda S^{-1}  \ 
\ee
(for invertible $F'$ this sufficient condition is also necessary). Therefore, 
\be  \label{Lam-Phi}
\Lambda  \left(I + h_n  A \Phi_1 S \right)  =   h_n  \Phi_1 S  \   .
\ee
Hence, \rf{lam} follows provided that \ $I + h_n  A \Phi_1 S$ \ is invertible. 
\end{Proof}

Therefore any discrete gradient scheme has a  locally exact modification of the form \rf{GR-LAM} provided that the right-hand side of \rf{lam} exists. This modification is unique if  $F'$ is invertible. By \rf{fi1} $\varphi_1 (h_n F')$ is invertible for sufficiently small $h_n$ and  $\varphi_1 (0) = I$. Hence,   $\left(I + h_n  A \Phi_1 S \right)^{-1}$ exists for sufficiently small $h_n$ and 
\be
  \lim_{h_n \rightarrow 0}  \frac{1}{h_n}  \Lambda = S \ . 
\ee 
\ods 
It turns out that $\Lambda$ given by   \rf{lam} is skew-symmetric,  $\Lambda^T = - \Lambda$. 
We point out that the skew symmetry of $\Lambda$  is  not assumed and is not obvious. In order to prove this fact we need the following lemma.  

\begin{lem}    \label{lem-g}
Suppose  $g (z)$ is an even analytic function ($g(-z)=g(z)$) and  $M$ is a matrix admitting  the factorization  \ $M = Q T$, where $T^T = T$,  $Q^T = - Q$ and $Q$ is invertible. Then
\be  \label{gT} 
   \left( g (M) \right)^T = Q^{-1} g (M) Q \ . 
\ee
\end{lem}  

\begin{Proof}  We represent  $g$ as a series
\be
   g (M) = \sum_{k=0}^{\infty}  a_k (Q T)^{2 k} .
\ee
Using assumed properties of $Q, T$ we have $(Q T)^T = - T Q$. Therefore 
\be
   \left( g (M) \right)^T =  \sum_{k=0}^{\infty}  a_k ( T Q )^{2 k}  = \sum_{k=0}^{\infty}  a_k Q^{-1} (Q T)^{2 k} Q \ . 
\ee
Hence, \rf{gT} follows. 
\end{Proof}

\ods

\begin{Th}  \label{Th-skew}
Matrix $\Lambda$ given by \rf{lam} is skew-symmetric. 
\end{Th}

\begin{Proof}  We assume that $h_n$ is sufficiently small so that  $\Phi_1$ and $\lambda$ are both invertible. Then, \rf{Lam-Phi} implies
\be
h_n  \Lambda^{-1} = S^{-1}  \left(   \Phi_1^{-1} + h_n S A \right)  .
\ee
We define  $R = A - B$. Then, taking into account $SA+SB= F'$, we get
\be
  S A  = \frac{1}{2}  F'  + \frac{1}{2} S R \ , \qquad  S B = \frac{1}{2}  F'  - \frac{1}{2} S  R \ . 
\ee
Hence,  
\be   \label{Lam-inv}
h_n  \Lambda^{-1} = \frac{1}{2} h_n R +  S^{-1}  g (F') \ , \qquad  g (F') =  \Phi_1^{-1} + \frac{1}{2} h_n F' \  .
\ee
Note that $g (z)$ is an analytic function on the disc  $h_n |z|< 2 \pi$, see \rf{fi1}. Using  Theorem~\ref{Th-AB}   we easily show skew-symmetry of $R$: 
\be  \label{RT} 
R^T = A^T - B^T = B-A=- R \ . 
\ee 
Moreover,  expressing $\varphi_1 (h_n F')$ is terms of $e^{h_n F'}$ (compare \rf{exp-fi}), we have
\be  \label{g-coth}
g (F') =  \left(  \varphi_1 (h_n F') \right)^{-1} + \frac{1}{2} h_n F' = \frac{1}{2} h_n F' \coth \left( \frac{1}{2} h_n F' \right) \ ,  
\ee
where $z \coth (z) \equiv z \cosh(z)/\sinh(z)$ is analytic for $|z| < \pi$. 
Hence   $g (- F') = g (F')$.  Moreover, 
$F' = S H_{\pmb{y} \pmb{y}}$, where $S^T = - S$ and $H_{\pmb{y} \pmb{y}}^T = H_{\pmb{y} \pmb{y}}$. Therefore, by Lemma~\ref{lem-g},  
\be 
  \left( g(F') \right)^T = S^{-1} g (F') S \ .
\ee
Then,  using $S^{-1} = S^T$ and  $S^2 = - I$, we obtain 
\be   \label{SGT}
  \left(  S^{-1}  g (F') \right)^T =  \left( g (F') \right)^T   S  = - S^{-1}  g (F') \ .
\ee
Taking into account  \rf{RT} and \rf{SGT}, we have from \rf{Lam-inv} that $\Lambda^{-1}$ is skew symmetric. Hence $\Lambda^T = - \Lambda$. 
\end{Proof}

Therefore $\Lambda$  from \rf{lam} satisfies  \rf{AS}, which means that the locally exact modification
defined by \rf{lam}  is also energy preserving (for any discrete gradient $\bar \nabla H$).

\section{Special cases}

Locally exact  modifications have simplest form in the case of symmetric discrete gradients. Then $A=B$ and  $R=0$. In this case \rf{Lam-inv} and \rf{g-coth} yield:  
\be  \label{lam-sym}
 \Lambda  = h_n  \tanhc \left( \frac{1}{2} h_n F' \right)  S  ,
\ee
where $\tanhc (z) := z^{-1} \tanh z$ for $z\neq 0$ and $\tanhc (0) := 1$. The function  $\tanhc (z)$ is analytic on the disc $\{ z\in \C: |z|<\pi/2 \}$, hence $\Lambda$ is well defined for sufficiently small $h_n$, compare  \rf{ser}.   
Note that formula \rf{lam-sym} is independent of the discrete gradient for all  symmetric  discrete gradients. 

Specializing results of the previous section to separable Hamiltonians of the form 
 \be  \label{H-sep}
  H (\pmb{x}, \pmb{p}) = T (\pmb{p}) + V (\pmb{x}) 
\ee
 and to any symmetric discrete gradient ${\bar \nabla}_s$,  we get the following useful theorem, compare \cite{Ci-locex},  Proposition 6.12.  We use notation $\tanc (z) := z^{-1} \tan (z)$ for $z\neq 0$ and $\tanc (0) := 1$ (functions $\tanc (z)$ and $\tanhc (z)$ are analytic in a neighbourhood of $z=0$). 

\ods

\begin{Th}   \label{Th-spec}
Numerical scheme 
\be \ba{l}  \label{grad-del}
\pmb{\delta}_n^{-1} \left( {\pmb x}_{n+1} - {\pmb x}_n \right) = {\bar \nabla}_s T (\pmb{p}_n, \pmb{p}_{n+1})    \\[2ex] \dis
(\pmb{\delta}_n^T)^{-1} \left( {\pmb p}_{n+1} - {\pmb p}_n \right) = - {\bar \nabla}_s V (\pmb{x}_n, \pmb{x}_{n+1})    
\ea \ee
 preserves exactly the energy integral (for any $h_n$-dependent invertible  matrix \ $\pmb{\delta}_n$ such that $\pmb{\delta_n} \rightarrow h_n I$ for $h_n \rightarrow 0$) and is locally exact for
\be  \label{delta_n}
   \pmb{\delta}_n = h_n  \tanc \frac{h_n \Omega_n}{2} \ , \qquad \Omega_n^2  = T_{\pmb{p} \pmb{p}} (\pmb{\bar p}_n) V_{\pmb{x} \pmb{x}} (\pmb{\bar x}_n)  ,
\ee
($\pmb{\delta}_n$ given by \rf{delta_n} is invertible for sufficiently small $h_n$, at least). 
\end{Th}

\begin{Proof}  The system \rf{grad-del} is equivalent to \rf{GR-LAM} if we take
 \be  \label{lam-del}
  \Lambda = \mm  0 & \pmb{\delta}_n \\  - \pmb{\delta}_n^T & 0 \ema .
\ee
Hence, by Lemma~\ref{lem-zachow}, the numerical scheme \rf{grad-del} is energy preserving. 

Since $\bar \nabla H$ is a symmetric discrete gradient  if $\Lambda$ is given by \rf{lam-sym}, then the scheme defined by \rf{grad-del}  is locally exact.   We have
  $F = (T_{\pmb{p}},  - V_{\pmb{x}})^T$ \ and  
\be
 F' = \mm  0 &  T_{\pmb{p} \pmb{p}} \\ 
- V_{\pmb{x} \pmb{x}} & 0 \ema  , \qquad (F')^2 = - \mm T_{\pmb{p} \pmb{p}} V_{\pmb{x} \pmb{x}} & 0 \\ 0 & V_{\pmb{x} \pmb{x}} T_{\pmb{p} \pmb{p}} \ema  ,
\ee
where all quantities are evaluated at $(\pmb{\bar x}_n, \pmb{\bar p}_n)$. 
Denoting  $\Omega_n^2 = T_{\pmb{p} \pmb{p}} V_{\pmb{x} \pmb{x}}$ we have:  
\be  \label{Fkwa}
 (F')^2 = - \mm \Omega_n^2 & 0 \\ 0 & (\Omega_n^T)^2 \ema  .
\ee
Comparing Taylor series of $\tanhc (z)$ and $\tanc(z)$ we see that  $\tanhc(A)=\tanc(B)$ for any matrices satisfying $B^2 = - A^2$. Therefore,  formula \rf{Fkwa} implies
\be
  \tanhc \frac{h_n F'}{2} = \mm   \tanc  \left( \frac{1}{2} h_n \Omega_n \right) & 0  \\ 0 &  \tanc  \left( \frac{1}{2} h_n \Omega_n^T \right)  \ema .
\ee
Therefore, if we define $\pmb{\delta}_n$  by \rf{delta_n}, then the formula  \rf{lam-del}  yields \rf{lam-sym}, which means that the scheme given by \rf{grad-del} is locally exact.  
\end{Proof}

In the case when $\pmb{x}$ and $\pmb{p}$ are scalars ($x$ and $p$), then  $(F')^2$ is  proportional to the unit matrix. Therefore, an energy-preserving locally exact scheme of the form 
\be \ba{l} \dis
\frac{x_{n+1} - x_n}{\delta_n} = \frac{\Delta H}{\Delta p}  \ , \qquad \frac{p_{n+1} - p_n}{\delta_n} = - \frac{\Delta H}{\Delta x}  \ , 
\ea \ee
exists for any Hamiltonian and any symmetric discrete gradient, compare \cite{CR-BIT}. It is enough to take  
\be  \label{delta-omH} 
  \delta_n = h_n  \tanc \frac{h_n \omega_n}{2} \ , \qquad \omega_n = \sqrt{ H_{xx} H_{pp} - H_{xp}^2 } \ , 
\ee
where $\omega_n$ is evaluated at $\bar x_n, \bar p_n$. 

We point out that the formula \rf{delta-omH} implies some limitations on $h_n$ in the case $\omega_n \in \R$. Certainly we have to require $h_n  \omega_n \neq  \pi + 2 \pi M$ ($M \in \N$), or even $h_n \omega_n <  \pi$.  The last inequality is not very restrictive because it means that $h_n < \frac{1}{2} T_n$, where $T_n = 2\pi/\omega_n$ is a corresponding period, see \cite{CR-PRE}.

The case \ $H (x, p) =  \frac{1}{2} p^2 + V (x)$ \ was  considered in previous papers \cite{CR-PRE,CR-BIT,CR-long}, where one can find results of many numerical experiments. 
Assuming \ $\bar x_n = x_n$, $\bar p_n = p_n$  \ we obtain a scheme of 3rd order (GR-LEX), while \ $\bar x_n = \frac{1}{2} \left( x_n + x_{n+1} \right)$, $\bar p_n = \frac{1}{2} \left( p_n + p_{n+1} \right)$ \ yields a scheme of 4th order (GR-SLEX). In the case of one degree of freedom the discrete gradient scheme without modifications is of second order.  Numerical experiments have shown that the accurcy of GR-LEX and GR-SLEX is higher by several orders of magnitude when compared with the standard discrete gradient method (while the computational cost is higher at most by several times, usually much less) \cite{CR-PRE,CR-BIT}.  We point out, however, that in the multidimensional case  the order of locally exact modifications is  usually not greater than 2, and can be higher only for exceptional  discrete gradients.

\begin{figure}
\caption{Error of numerical solutions  for a circular orbit ($R=0.1$, exact period $T\approx6.28$) in the potential $V (r) = 0.5 r^2  - 0.01 r^4$ as a function of $t$.  GR: dark line ($h=0.5$),  GR-SYM: black line ($h=0.625$),   GR-LEX: gray line ($h=0.766$), GR-SLEX: light gray line ($h=1.063$). }
 \label{fig-0.1-0.5}  \par   \ods
\includegraphics[width=0.9\textwidth]{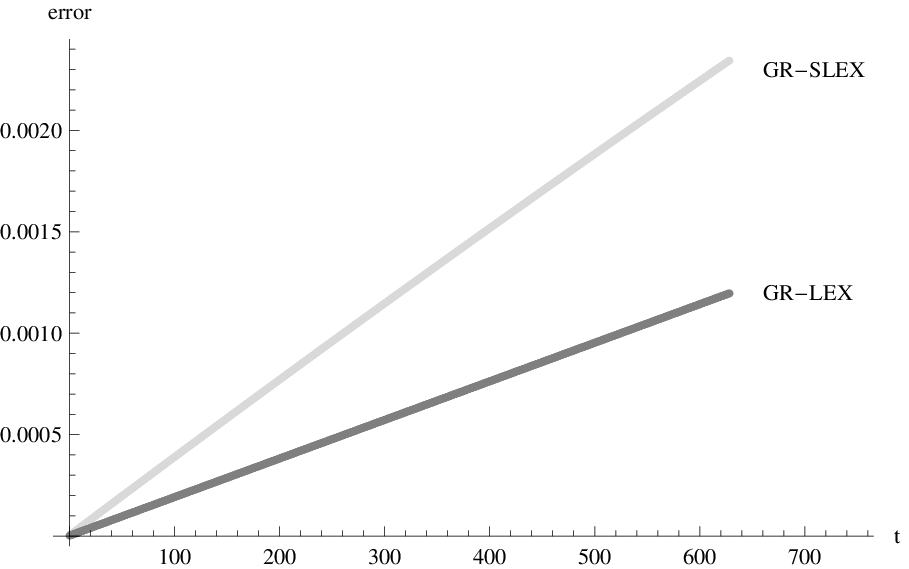} \par
\includegraphics[width=0.9\textwidth]{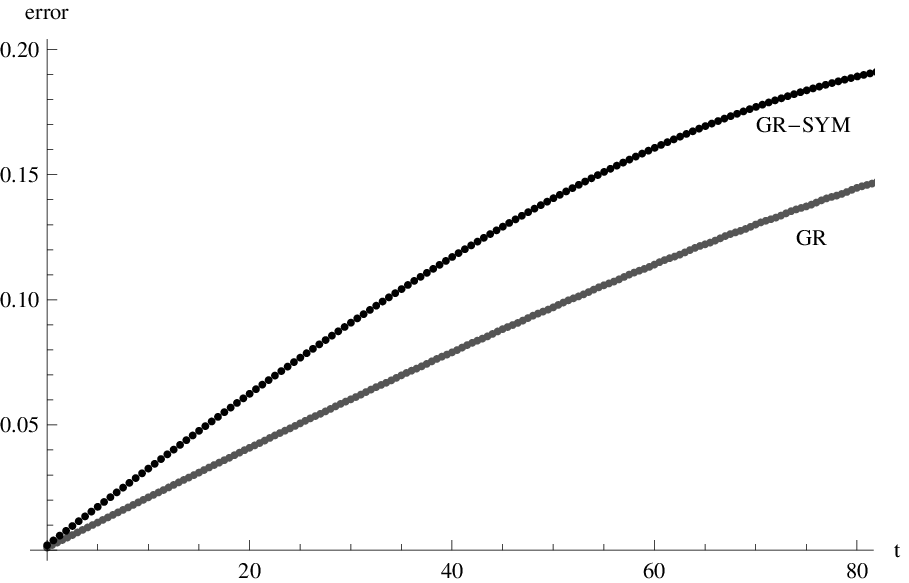} \par
\end{figure}

\begin{figure}
\caption{Error of numerical solutions   for a circular orbit ($R=1.0$, exact period $T\approx6.41$) in the potential $V (r) = 0.5 r^2  - 0.01 r^4$ as a function of $t$.   GR: dark line ($h=0.05$),  GR-SYM: black line ($h=0.067$),   GR-LEX: gray line ($h=0.094$), GR-SLEX: light gray line ($h=0.154$). }
 \label{fig-1-0.05}  \par   \ods
\includegraphics[width=\textwidth]{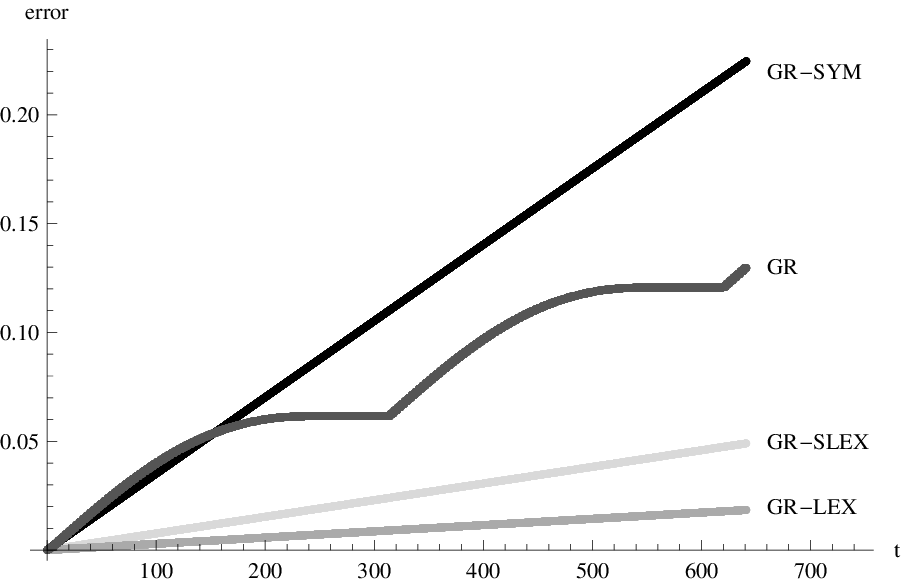} \par
\end{figure}

\begin{figure}
\caption{Error of numerical solutions (at $t = 641$) for a circular orbit ($R=1.0$, exact period $T\approx6.41$) in the potential $V (r) = 0.5 r^2  - 0.01 r^4$ as a function of $t$.  GR: dark line ($h=0.5$),  GR-SYM: black line ($h=0.625$),   GR-LEX: gray line ($h=0.766$), GR-SLEX: light gray line ($h=1.063$). }
 \label{fig-1-0.5}  \ods
\includegraphics[width=\textwidth]{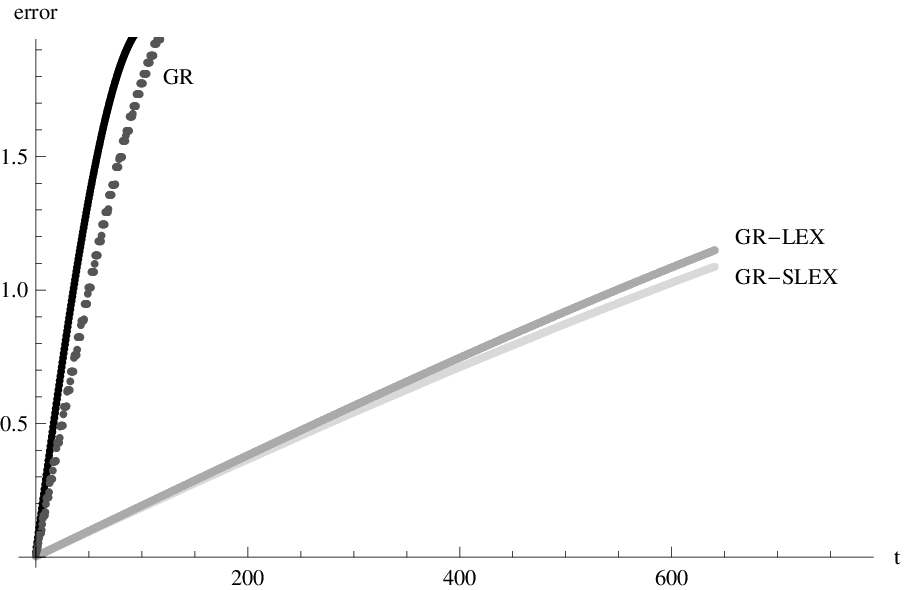} \par
\end{figure}

\begin{figure}
\caption{Error of numerical solutions   for a circular orbit ($R=3.0$, exact period $T\approx 7.85$) in the potential $V (r) = 0.5 r^2  - 0.01 r^4$ as a function of $t$.   GR: dark line ($h=0.5$),  GR-SYM: black line ($h=0.627$),   GR-LEX: gray line ($h=0.768$), GR-SLEX: light gray line ($h=1.066$). }
 \label{fig-3-0.5}  \par   \ods
\includegraphics[width=\textwidth]{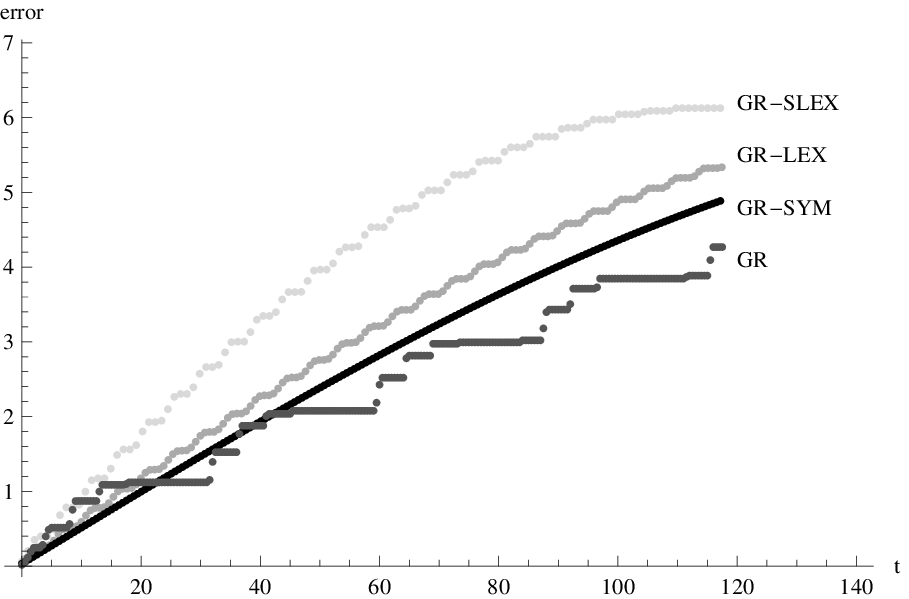} \par
\end{figure}

 In this Letter we present few results of numerical experiments for a two-dimensional anharmonic oscillator 
\be
H (\pmb{x}, \pmb{p}) = \frac{1}{2} |\pmb{p}|^2 + V (|\pmb{x}|) \ , \quad  V (|\pmb{x}|) = \frac{1}{2} |\pmb{x}|^2 - \frac{1}{100} |\pmb{x}|^4 .
\ee
see Figs.~\ref{fig-0.1-0.5}, \ref{fig-1-0.05}, \ref{fig-1-0.5} and \ref{fig-3-0.5}. 
We consider circular orbits (of radius $R$), when the exact solution can be easily found (in this potential  the radius of a circular orbit  have to be smaller than $5$).  We compare  the 
 coordinate increment discrete gradient \rf{grad-incre} (denoted by GR),  its symmetric modification \rf{grad-sym} (denoted by GR-SYM), and locally exact modifications \rf{grad-del}:  GR-LEX ($\pmb{\bar x} = \pmb{ x}_n$) and  GR-SLEX ($\pmb{\bar x} = \frac{1}{2} ( \pmb{\bar x}_n +\pmb{x}_{n+1}$)).  In this case GR-SYM, GR-LEX and GR-SLEX are of  the second order, while GR is only of the first order. 

Locally exact modifications are more expensive. The cost was estimated by the number of function evaluations. The average number of function  evaluations per step depends on $h$ and on the method. For instance, in the case of  GR we have 85 evaluation per step for $h=0.05$ and 160 evaluations per step for $h=0.5$. In the case of GR-SLEX we have 262 and 341 evaluations per step, respectively. In numerical experiments we use different time steps in order to get  the same computational costs. 

Fig.~\ref{fig-0.1-0.5} shows that for $R=0.1$  locally exact modifications are more accurate than standard discrete gradient schemes by about 3 orders of magnitude. We had to divide this figure into two parts (with different scales on axes).   Figs.~\ref{fig-1-0.05} and \ref{fig-1-0.5}  concern the case $R=1$. Locally exact schemes are more accurate by several times. Note that both figures are quite similar although time steps are much smaller at  Fig.~\ref{fig-1-0.05}.  The efficiency of locally exact modifications decreases with increasing $R$. For  $R=3$  all four considered schemes have similar accuracy but  surprisingly GR is the best, see Fig.~\ref{fig-3-0.5}.  In the case of  larger $R$  our modifications are less accurate, especially when compared with  GR-SYM. Therefore, we can conclude that locally exact schemes are very accurate in a neighbourhood (not very small, in fact) of the stable equilibrium.

\section{Concluding remarks}

We presented a construction of new numerical schemes based on the notion of local exactness. This notion has been    known (under different names)  for  almost fifty years. The original application was confined to  exact discretization of linearized equations \cite{Pope}. Our approach has two new features. First, we modify a given numerical scheme in a  locally exact way. Second, we try to preserve geometric properties of the original numerical scheme. This task is not trivial. In this paper we present one successful  application: locally exact modifcations of discrete gradient methods for canonical Hamilton equations. 
We constructed a locally exact integrator (a modification of the discrete gradient scheme) which preserves exactly the energy integral. 

In the case of one degree of freedom the proposed modification, although more expensive (by only a few  percent), turns out to be more accurate by as much as  8 orders of magnitude (in the case of small oscillations around the stable equilibrium) in comparison to the standard discrete gradient scheme \cite{CR-PRE,CR-BIT}. In multidimensional cases the relative cost of our algorithm is higher, but still we hope that our method will be of advantage.  In general, the accuracy of locally exact algorithms is very high (much higher than their order suggests) in the neighbourhood of  stable equilibria.  
Modifications proposed in this Letter contain exponentials of variable matrices. Similar time-consuming evaluations are characteristic for all exponential integrators. Effective methods of computing matrix exponentials, recently developed in this context \cite{HoL1,NW},  may decrease the computational cost of our algorithms.  

 Another advantage of the proposed approach is the natural possibility of using a  variable time step. Unlike symplectic methods (which work mostly for constant time step) discrete gradient methods admit conservative modifications with variable time step. Therefore, one may easily implement any variable step method in order to obtain further improvement. 
\ods

\no {\it Acknowledgements.} 
This work is partly supported by the National Science Centre (NCN) grant no. 2011/01/B/ST1/05137.  I am grateful to an anonymous referee for detailed  comments which helped to improve this Letter.

\end{document}